# Cristiano Ronaldo or Lionel Messi, who is more consistent in scoring goals? The evidence from CFM exploratory analysis

Samsul Anwar[1,*] , Siti Munawarah[2], Radhiah Radhiah[3]

[1,2] Universitas Syiah Kuala, Department of Statistics, Banda Aceh, Indonesia
[3] Universitas Syiah Kuala, Department of Mathematics, Banda Aceh, Indonesia

**Abstract** – The rivalry between two football superstars Cristiano Ronaldo and Lionel Messi has always been a subject of extensive discussion. This study aimed to compare the level of consistency between the two players in scoring goals through 6 ways: right-footed kicks, left-footed kicks, penalty kicks, direct free kicks, long-range kicks, and headers. The data analyzed was the duration of time (minutes) each player took to score a goal in every match they played. The data was obtained from a football website called Transfermarkt.com. Competing Failure Modes (CFM) was used to measure the reliability of the two players in scoring goals based on those various ways. The results of CFM exploratory analysis showed that Ronaldo and Messi had the same level of consistency in scoring goals for more than 17 years of their professional football career. Both have been among most talented players in the modern football era with individual and team achievements that are far above other footballers around the world.

**Résumé - Cristiano Ronaldo ou Lionel Messi, qui est le plus régulier pour marquer des buts? Une preuve tirée des résultats de l'analyse exploratoire CFM.** La rivalité entre les deux superstars du football Cristiano Ronaldo et Lionel Messi a toujours fait l'objet de nombreuses discussions. Cette étude visait à comparer le niveau de cohérence entre les deux joueurs dans la notation des buts de 6 manières: coups de pied droit, coups de pied gauche, tirs au but, coups francs directs, coups de pied lointains et coups de tête. Les données analysées étaient la durée (minutes) que chaque joueur mettait pour marquer un but dans chaque match qu'il a disputé. Les données ont été obtenues à partir d'un site web de football appelé Transfermarkt.com. Des Competing Failure Modes (CFM) ont été utilisés pour mesurer la fiabilité des deux joueurs à marquer des buts en fonction de ces différentes manières. Les résultats de l'analyse exploratoire du CFM ont montré que Ronaldo et Messi avaient même niveau de régularité pour ce qui est de marquer de buts pendant plus de 17 ans de leur carrière de footballeur professionnel. Tous deux comptent parmi les joueurs les plus talentueux de l'ère du football moderne, avec des réalisations individuelles et collectives bien supérieures à celles des autres footballeurs du monde.

**Mots clés:** modes de défaillance concurrents, football, fiabilité des buts marqués, transfermarkt

---

*Corresponding author: samsul.anwar@usk.ac.id

# 1 Introduction

Football is the world's most popular sport since it is played in almost every country around the world (Busey & Waring, 2012). The world football federation, Fédération Internationale de Football Association (FIFA), estimates that there are currently around 250 million football players and more than 1.3 billion people interested in football (Giulianotti et al., 2021).

Everyone in football starting from general managers and coaching staff to fans and media are interested in evaluating player's performance (Pelechrinis & Winston, 2021). Historically, football has had many talented players who serve as inspirational models for young footballers and are still discussed and followed today. A football legend is a player who managed to carve his name in the football history books, not only based on his personal success records, but also for a team he played on. In the modern era of football in the 21st century, there are two players who are always associated with each other, namely Cristiano Ronaldo and Lionel Messi. Apart from Ronaldo and Messi, there are indeed several other famous modern football players such as Zlatan Ibrahimović, Neymar, Luis Suárez, Mohamed Salah, Robert Lewandowski, Karim Benzema and others. But all these players have been rarely in the media spotlight like Ronaldo and Messi since these two players played on the well regarded Spanish La Liga clubs, namely Real Madrid C.F. and FC Barcelona in 2009.

According to Castañer et al. (2017), both are considered the best football players ever. Ronaldo and Messi were listed as the most footballers to win the FIFA Best Player award or known as the FIFA Ballon d'Or. From 2008 to 2023, Messi won 8 times and Ronaldo 5 times. In addition, these two football superstars have made numerous achievements. Based on trophies at club level, Messi has won 38 trophies with FC Barcelona, Paris Saint-Germain F.C. and Inter Miami CF while Ronaldo has won 31 trophies throughout his career at several clubs he has played for, including Sporting CP, Manchester United F.C., Real Madrid C.F., Juventus F.C. and Al Nassr FC. Furthermore, for trophies with the national team, Ronaldo succeeded in bringing Portugal to win two trophies, namely the UEFA Euro 2016 Cup and the 2019 UEFA Nations League. On the other hand, Messi managed to bring Argentina to win the Copa América (2021 and 2024) and the 2022 World Cup. There are many other individual and team achievements of the two players that are not mentioned here. We can easily find it in their biographies which are published in various existing literature.

The debate over who is greater between Ronaldo and Messi will never stop until both hang up their boots. According to Shergold (2016), the debate about these two football superstars will never end. These two figures have been very attractive to the football world community. Tiago et al. (2016) claimed that for most athletes including football players, social media is a powerful tool to take advantage of their time in the spotlight. Fans of both will never tire of claiming their respective idols as the best one. A study by Kassing (2020) showed that fans used internet sports memes to belittle of their rival opponents and idolize their own players.

Based on the data collected in a sport, researchers are able to perform quantitative analysis on the performance of players and teams (Rahimian & Toka, 2022). Consistency in showing the best performance in every match is an important indicator in assessing the career path of a professional football player. In order to prove who is more dominant between these two most famous players, several scientific studies have been conducted. A research by Castañer et al. (2016) using a polar coordinate analysis approach shows that Messi is a player who is more dominant in using his left foot. Messi has excellent laterality skills (left-right brain balance) as a football player. Furthermore, Shergold (2016), who conducted research using descriptive comparative analysis when both played in the Spanish La Liga since the 2009-10 season stated that Messi managed to score 270 goals in 252 matches, played for 21,218 minutes and shot 953 times. Ronaldo managed to score 270 goals in 247 matches, played for 21,206 minutes and shot 1,318 times. The study showed that Messi had a shot conversion rate of 28.77%, while Ronaldo had 20.03%. Apart from having a high level of accuracy, both also have uniqueness in motor skills that are different from other football players.

Magee (2025) showed that having more attacking players on the pitch will result a higher rate of scoring goals, signifying the importance of the forwards in helping the team to secure the game. Various studies from several competitions denoted around 40 to 50% of goals were scored by attacking players, indicating a consistency in scoring goals is important and represent the ability of a footballer, especially forwards, in adapting the change that occurs from a season to season (Acar et al., 2009; M. Mitrotasios & Armatas, 2014; M. Mitrotasios et al., 2006). Where the opponents they might have faced were likely also changed in terms of their player composition, strategy implemented, level of play, and so on. Accordingly, it is important to measure the level of consistency in scoring goals over a long period of time as an indicator to compare the professional career paths of Ronaldo and Messi as forwards. Unfortunately, the previous studies have not been able to answer who had more consistent performance between Cristiano Ronaldo and Lionel Messi during their entire careers as professional football players. The comparison carried out in the first study focused more on comparing the physical abilities between the two players, while the comparison in the second study was more directed at the performance between the two players through analysis using aggregate data especially when both were playing in La Liga, so that the conclusions obtained are not very accurate and only apply to that time period.

To scientifically prove who had more consistent performance between Ronaldo and Messi, this study employed an alternative approach based on time intervals to score a goal. To the best of our knowledge, this is the first study applying this approach in measuring reliability of a football player in scoring goals. Apart from the number and duration of time needed to score a goal, this study also considered different ways of scoring goals. In football, there are several ways to score a goal, through penalty kick, header, direct free kick, long-range kick, right-footed kick and left-footed kick. Each of these ways of scoring goals has a different level of difficulty.

Therefore, the objective of the current study was to compare the goal scoring reliability between Cristiano Ronaldo and Lionel Messi to find out who had more consistent performance between these two recognizable players during their almost entire career as professional football players. This exploratory study has added some objective assessment indicators for those who may require comparing performance of professional football players, especially between Ronaldo and Messi.

## 2 Methods

### 2.1 Data, methods and variables

According to Collett (2015), survival analysis is statistical analysis of duration time data from the beginning of the study until certain events occur. In this study, the event was defined as goals scored by Cristiano Ronaldo and Lionel Messi in every match they played. The data examined in this study was the duration of time (in minutes) each player took to score a goal calculated since Ronaldo and Messi began to be deployed to play on the pitch. In general, survival data can be divided into 2, namely censored data (incomplete) and uncensored data (complete data). In this study, when the player successfully scored a goal during the game, the time duration data was categorized as uncensored data. Conversely, if the player did not score any goal, the data was categorized as censored data.

Commonly, an event that occurs in the object of research in survival analysis is only caused by one mode. However, in practice, the same event may occur due to several possible modes. Competing Failure Modes (CFM) analysis is one of the analytical methods used to analyze this such event (ReliaSoft, 2015). Just like other methods in survival analysis, CFM could also be applied to other various fields. The current study tried to apply this analysis in the field of sports, especially football. In a football game, a goal can be scored in several ways (modes) such as through a right-footed kick, left-footed kick, penalty kick, free kick, long-range kick, and head header. CFM analysis was used to measure the reliability of Ronaldo and Messi in scoring goals by considering those different ways of goals scored. Simply put, CFM analysis in this study measured the combined reliability of all these ways of scoring goals. In other words, the probability of a player still not scoring any goal after he played for a specific time period. Accordingly, a player with lower reliability value was considered better in scoring goals than their counterpart.

To estimate the reliability value of each mode of the event (a goal scored in this study), the non-parametric method Kaplan-Meier was used. We considered employing a non-parametric method because it did not require certain probability distribution assumptions so that the comparison of the duration of time taken to score a goal between the two players was done without a need to distinguish the probability distribution of the data. Through the CFM exploratory analysis, it became known who was more consistent in scoring goals between Ronaldo and Messi.

The data on the duration of time taken by the two players to score goals used in this study was the secondary data extracted from a German-based football website called Transfermarkt (*https://www.transfermarkt.com/*). The website has provided all information about football matches around the world including scores, results, statistics, transfer news, and fixtures and recorded all game statistics both for teams and players for every season. There were more than 60.000 games from many seasons in all major football competitions. The website (at the time of conducting this study) was available in fourteen languages and the player database contained details of more than 800,000 professional footballers. Transfermarkt obtained data about these players from more than 1 million of their registered users around the world who kept track of how many minutes each player spends on the pitch, medical history, contract duration, and many more. With more than one billion page views per month, Transfermarkt has become one of the world's largest football websites. The website has become the unofficial player appraiser of the football world. Many top clubs and international media (e.g. ESPN, L'Équipe, The Guardian, and La Gazzetta dello Sport) cite Transfermarkt as a source (Keppel & Claessens, 2020).

The data period used in this study spans from the 2002-03 season to the 2020-21 season. This period was chosen because Ronaldo began his professional career as a football player in the 2002-03 season, while Messi made his debut as a professional football player two seasons later. In analyzing these two recognizable footballers' careers, Rodriguez and Villablanca (2024) suggested analyzing the data since the beginning of their professional career just like we did in this study. In addition, to reduce potential biases due to this different starting time career, we employed a survival analysis approach that focused more on the duration of time an event occurs rather than the number of samples analyzed.

The variables analyzed in this study consisted of the duration of time taken to score a goal ($T$), status of censoring data ($d$) and the ways of goals scored ($X$) as shown in Table 1.

### 2.2 Research procedures

Data analysis in this study was carried out separately between Cristiano Ronaldo and Lionel Messi using R software version 4.0.3 with the survival package. This study was conducted using a nine-step procedure. The first one was collecting research data, namely game statistics of Ronaldo and Messi in all competitions starting from the 2002-03 season to the 2020-21 season. The data consisted of the time duration needed to score a goal and other variables related to this study. The second step was separating the data for both players based on the way the goals were scored (penalty kick, head header, direct free kick, long-range kick, right-footed kick and left-footed kick). The third one was conducting descriptive statistics analysis to obtain an overview of the data characteristics of the time duration it took for both players to score a goal.

**Table 1.** Study Variables

| Variables | Description | Scale | Category |
|---|---|---|---|
| $T$ | Duration of time taken to score a goal (in minutes) | Ratio | - |
| $d$ | Status of censoring data | Nominal | 0 : Censored<br>1 : Uncensored |
| $X$ | Ways of goals scored | Nominal | 1 : Penalty Kick<br>2 : Head Header<br>3 : Direct Free Kick<br>4 : Long Distance Kick<br>5 : Right-Footed Kick<br>6 : Left-Footed Kick |

To quantify the performance, the player who had more goals per minute of game played was awarded 1 point, and the total points earned by each player were compared for the first half (minutes 1 - 45), second half (minutes 46 - 90 minutes) and extra time (minutes 91 - 120).

The fourth one was employing non-parametric Kaplan-Meier method to estimate the reliability value of each way of scoring goals for both players. The Kaplan-Meier reliability value ($\hat{R}(t)$) was calculated using Equation (1).

$$\hat{R}(t) = \prod_{j: t_j \leq t} \left(1 - \frac{r_j}{n_j}\right) \quad (1)$$

where $r_j$ is the number of events (i.e. goals) that happened at time $t_j$ (time when at least one event happened) and $n_j$ is number of observation that have not yet had an event up to time $t_j$ (ReliaSoft, 2015).

The fifth step was performing Log Rank tests to evaluate the difference in Kaplan-Meier reliability curves between Ronaldo and Messi in each way of scoring goals. The sixth one was estimating the overall reliability value ($\hat{R}_{CFM}(t)$) and its 95% confidence intervals using CFM approach which was the product of reliability of each ways of scoring goals. The overall reliability value is defined as Equation (2).

$$\hat{R}_{CFM}(t) = \prod_i \hat{R}_i(t), i = 1, 2, 3, 4, 5, 6 \quad (2)$$

Note that if a modality of scoring goal was limited, then CFM only considered the way of scoring goals with a sufficient number of goals. The estimate 95% confidence intervals of each ways of scoring goals is defined as Equation (3).

$$\left[ \frac{\hat{R}(t)}{\hat{R}(t)+(1-\hat{R}(t)) \times e^{1.96 \times \frac{\sqrt{\widehat{Var}(\hat{R}(t))}}{\hat{R}(t) \times (1-\hat{R}(t))}}}, \frac{\hat{R}(t)}{\hat{R}(t)+(1-\hat{R}(t))/e^{1.96 \times \frac{\sqrt{\widehat{Var}(\hat{R}(t))}}{\hat{R}(t) \times (1-\hat{R}(t))}}} \right] \quad (3)$$

where $\widehat{Var}(\hat{R}(t))$ is the estimate variance of the reliability values (ReliaSoft, 2015). If the confidence interval between two or more compared groups overlap, it can be concluded that the differences between the groups were not significant (Cumming and Finch, 2005). The seventh step was plotting and comparing the overall reliability values in scoring goals ($\hat{R}_{CFM}(t)$) of Ronaldo and Messi based on CFM analysis. The eight one was interpreting the results of the study and discussing the comparison of goal scoring reliability between Ronaldo and Messi. Finally, the last step was drawing conclusions based on the results and discussions that were carried out before.

## 3 Results

### 3.1 Overview of research data

Table 2 provides data characteristics and goals distribution of Cristiano Ronaldo and Lionel Messi based on censoring status and the ways of goal scored. Based on Table 2, it can be inferred that in every match he played, Ronaldo managed to score 0.72 goals, while Messi was slightly higher at 0.80 goals/match. Moreover, Ronaldo managed to score more than 100 goals through 4 different ways namely through right-footed kicks, left-footed kicks, penalties and head headers. On the other hand, left-footed kicks were the only way Messi scored more than 100 goals. Note that the number of games played differs from the total data set analyzed, as a player might not score any goals, score one goal, two goals and so on for every game they played.

In addition to descriptive statistics based on censoring status and the way goals were scored, the comparison of goal distribution between the two players is also presented in more detail for each minute of the games they played. Figure 1 describes the number of goals scored by both players in every minute of the game including the extra time.

5**Table 2.** Data characteristics and goals distribution

| Variables | Cristiano Ronaldo | | Lionel Messi | |
|---|---|---|---|---|
| | Frequency | Percentage | Frequency | Percentage |
| Number of games Played | 1,089 | 100.00 | 941 | 100.00 |
| Status of censoring data | | | | |
|     Censored (did not score a goal) | 564 | 41.75 | 449 | 37.32 |
|     Uncensored (scored a goal) | 787 | 58.25 | 754 | 62.68 |
| Total dataset analyzed | 1,351 | 100.00 | 1,203 | 100.00 |
| Ways of goals scored | | | | |
|     Penalty Kick | 137 | 17.41 | 99 | 13.13 |
|     Head Header | 136 | 17.28 | 28 | 3.71 |
|     Direct Free Kick | 57 | 7.24 | 57 | 7.56 |
|     Long Distance Kick | 11 | 1.40 | 1 | 0.13 |
|     Right-Footed Kick | 303 | 38.50 | 92 | 12.20 |
|     Left-Footed Kick | 143 | 18.17 | 477 | 63.26 |
| Number of games scoring at least 1 Goal | 525 | 48.21 | 492 | 52.28 |

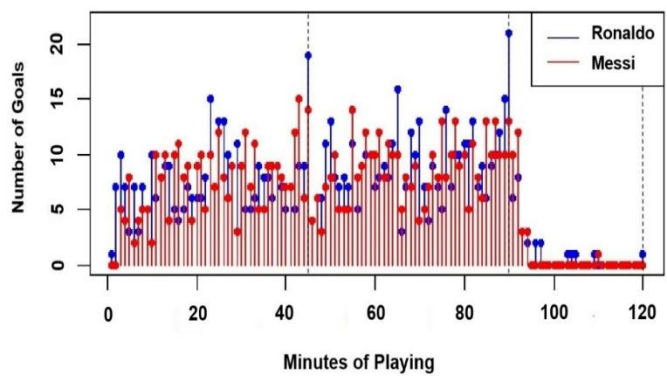

**Figure 1.** Distribution numbers of goals scored based on minute of playing

A comparison of goal scoring reliability between Ronaldo and Messi can be made by looking at who had more goals in every minute of the games they played. Total points earned by both players in each minute of playing are presented in Table 3.

**Table 3.** Total points represent the superiority in the number of goals between Cristiano Ronaldo and Lionel Messi

| Superior Player | Minutes of Playing | | | | | |
|---|---|---|---|---|---|---|
| | 1 - 45 | | 46 - 90 | | 91 - 120 | |
| | Points | % | Points | % | Points | % |
| Cristiano Ronaldo | 19 | 42,22 | 22 | 48,89 | 7 | 23,33 |
| Lionel Messi | 19 | 42,22 | 19 | 42,22 | 4 | 13,33 |
| Draw | 7 | 15,56 | 4 | 8,89 | 19 | 63,33 |
| Total | 45 | 100,00 | 45 | 100,00 | 30 | 100,00 |

Based on Table 3, it is known that overall Ronaldo managed to gain 48 points, or 6 points more than Messi.

**3.2 Reliability comparison in scoring goals between Ronaldo and Messi**

Figure 2 presents the Kaplan Meier reliability curves of scoring reliability for Cristiano Ronaldo and Lionel Messi based on 6 ways of goals scored.

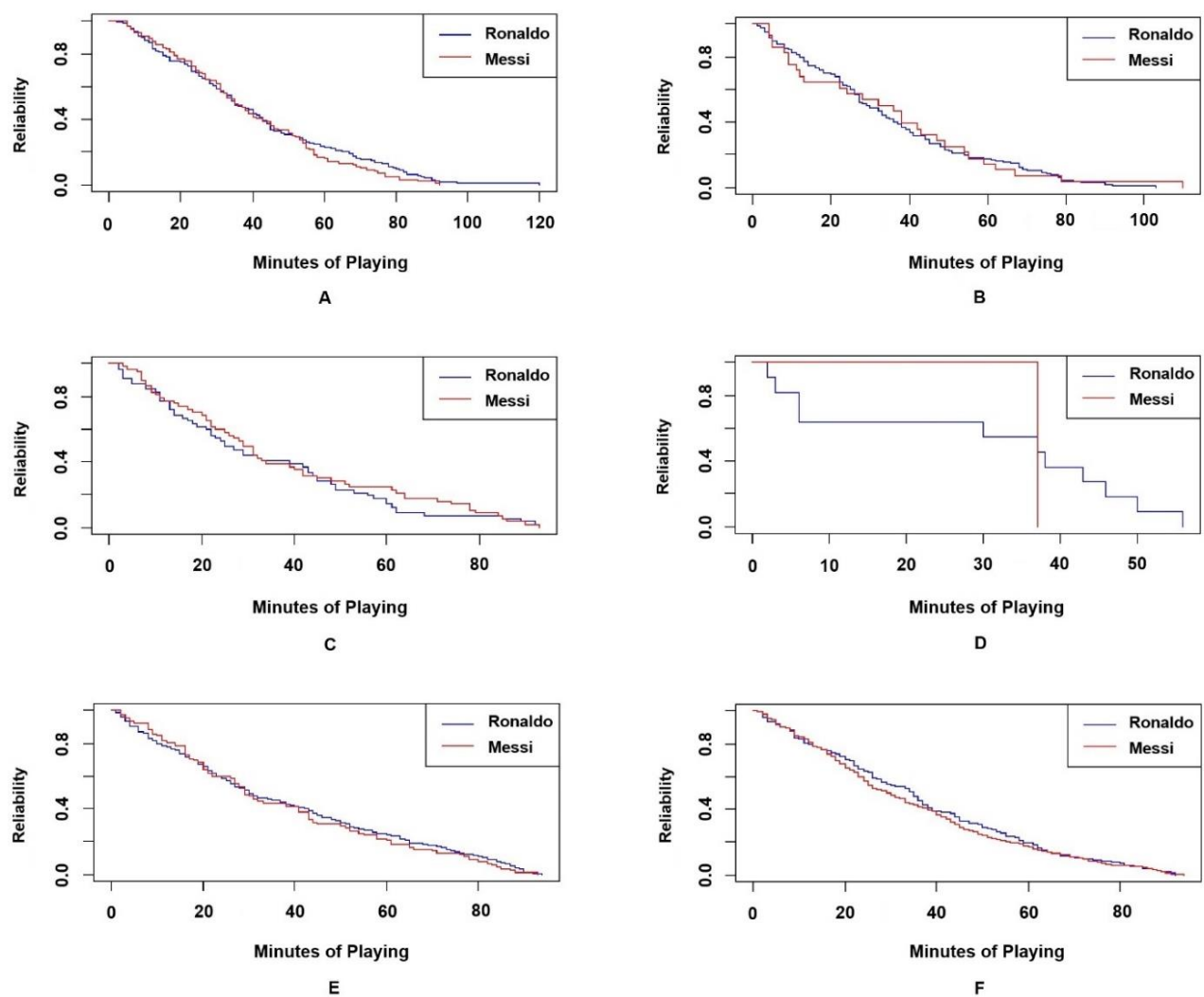

**Figure 2.** Kaplan Meier reliability curves for penalty kick (A), head header (B), direct free kick (C), long-range kick (D), right-footed kick (E) and left-footed kick (F)

While the Log rank test results are presented in Table 4.

**Table 4.** Log rank test on Kaplan-Meier reliability curves of Cristiano Ronaldo and Lionel Messi

| Ways of Goals Scored | Chi-square statistics | p-value |
|---|---|---|
| Penalty kick | 0.90 | 0.300 |
| Head header | 0.00 | 0.800 |
| Direct free kick | 0.40 | 0.500 |
| Long-range kick | - | - |
| Right-footed kick | 0.50 | 0.500 |
| Left-footed kick | 0.50 | 0.500 |

Based on Figure 2.A, it is known that the reliability curves of Ronaldo and Messi in scoring goals through Penalty kicks overlap with each other starting from the first minute of the game they played to the 50s. Although the reliability curves look slightly different especially in the 50s to 90th minute time interval of the games they played, the log rank test proved that the difference was not significant since p-value of the test was higher than the significance level of 0.05 (Table 4).

Figure 2.B shows that Ronaldo and Messi's reliability curves for goals scored using headers intersect each other in many observation times, starting from the first minute of the game they played to the 110th minute (extra time). This result was reinforced through Log Rank testing which showed that the reliability curves of the two players were not significantly different.

Although the reliability curves for the direct free kick in Figure 2.C appeared to be separate, further testing proved that the difference was not significant. On the other hand, the reliability curve for goals scored from long-range kicks (Figure 2.D) looks slightly different from the others. The observations for both players did not reach the 90th minute of the normal time. This was due to the fact that neither Ronaldo nor Messi scored any goals through long-range kicks above the 60th minute of the games they played. In addition, Messi's reliability curve looks very simple and only takes the form of a ladder. This was because the number of goals scored by Messi through long-range kicks was only 1 goal. Meanwhile, Ronaldo was able to score 11 goals through long-range kicks during his 17 years of career. Even so, the number of observations in this analysis was very small, especially for Messi and did not allow further analysis through the Log Rank test.

Ronaldo and Messi's reliability curves for goals scored using right-footed kicks (Figure 2.E) appear to overlap almost throughout the observation time, starting from the first to the final minute of the game they played. This was in line with the results of the Log Rank test (p-value more than 0.05). Finally, although the reliability curves for left-footed kicks of both key players seemed to be separate (especially in the span of 15 and 60th minutes), further testing proved that the reliability curves of the two players (Figure 2.F) were not significantly different.

It should be noted that, based on the results of the previous reliability analysis (Figure 2.D), it is known that the number of goals scored by both players through long-range kicks was limited. Therefore, the CFM analysis in this study only considered the reliability values of 5 ways of scoring goals, namely through penalty kick, head header, direct free kick, right-footed kick and left-footed kick. The CFM reliability value of each player is simply the multiplication of those five reliability values of the ways goals scored. Figure 3 presents the CFM reliability curves for Cristiano Ronaldo and Lionel Messi along with its 95% confidence intervals.

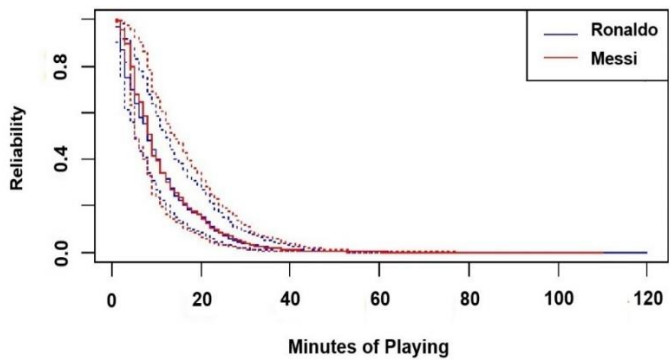

**Figure 3.** CFM reliability curves of Cristiano Ronaldo and Lionel Messi

The CFM reliability curves for the two players in Figure 3 are shown by different colors, blue for Cristiano Ronaldo and red for Lionel Messi. The continuous line shows the CFM reliability value, and the dashed line shows the 95% confidence interval. Based on Figure 3, it is known that Ronaldo and Messi's CFM reliability curves and its confidence intervals overlap with each other almost throughout the observation time, starting from the first minute of the games they played to the end of the extra time (120 minutes). In line with the Kaplan Meier reliability curves for each way of goals scored, the CFM reliability curves for the two players also did not differ significantly based on Cumming and Finch's confidence interval criteria.

## 4 Discussion

The objective of this study was to compare the level of reliability in scoring goals between Cristiano Ronaldo and Lionel Messi during most of their professional careers as footballers. The current study through CFM exploratory analysis showed that Cristiano Ronaldo and Lionel Messi had the same reliability in scoring goals in their golden era starting from the 2002-03 season to the 2020-21 season.

The recent study is in line with a study conducted by Anderson et al. (2020) who stated that voting for the best soccer player in the world (i.e. The Ballon d'Or) was tight, especially in 2016. While Messi received a higher percentage of the first place votes compared to Ronaldo (24.1% versus 23.1%), Ronaldo achieved higher average points compared to Messi on the second place votes (1.78 points versus 1.73 points). However, there are other studies that favored Messi more than Ronaldo or vice versa. Hussaini (2021) argued that Messi was the first and best player in the world, while Cristiano Ronaldo was the 6th best player in the world using the 2021 FIFA Ballon d'Or statistics. On the other hand, Salvador et al. (2022) stated that Ronaldo was in the first ranking while Messi in the second based on data of the 2019/2020 season. Interestingly, a study by Grbec et al. (2024) using longer periods of data showed that both were not even included in top ten ranking. In their study, Ronaldo was in the twelfth position, while Messi was in the top 14. The differentiation among the mentioned studies was due to the diversity of the seasons evaluated, methods employed and factors investigated since so far there was no standard approach to measure the performance of a footballer.

A study by Arora (2024) revealed differentiation of skills and roles of Ronaldo and Messi on the pitch. Messi has been well-known for his playmaking, vision, and dribbling prowess. His lightning-fast direction changes and illustrious left foot have been so dangerous for the opponents on the pitch. Messi created opportunities and scores goals by playing as an offensive midfielder or forward. On the other hand, Ronaldo has been known for his finishing, aerial ability, and physicality. His quickness, agility, and heading prowess has made him a versatile forward with an excellent goal scoring ability. While Messi was often praised for his natural skill and consistent play, Ronaldo was praised for his dedication, work ethic and versatility across a range of leagues.

Both Ronaldo and Messi have achieved unprecedented success, becoming global icons and setting numerous records. These individual and team achievements show that both were the most prominent players in the modern era of football, where the quality of both was quite equal but far above other players around the world. However, both had different characters in terms of their scoring styles where Ronaldo was more dominant in scoring goals with his right foot, while Messi with his left foot (Lama & Praveen, 2017). Yuwono and Rachman (2021) mentioned that the kicking power of a football player is significantly related to his training method and leg strength. Uniquely, Ronaldo has a special technique called *knuckling shot* in doing free kicks (Hong et al., 2012).

Although it looks easy, scoring goals through penalty kicks requires a strong mentality and high self-confidence (Mardhika

& Dimyati, 2015). Our study revealed that more than 17% of the goals scored by Ronaldo and 13% of the goals scored by Messi over their 17-year career came from penalty kicks. This proved that both have a good mentality and high confidence in executing penalty kicks. Furthermore, both were equally adept at scoring from direct free kicks. Around 7% of the total goals scored by both players over more than their 17-year professional careers came from direct free kicks. However, Ronaldo had a better accuracy rate in scoring goals through headers (Muminin, 2015). The data revealed that 17.28% of Ronaldo's goals were created by his head, while Messi only scored 3.71% of his total goals through headers.

Both players also had the highest number of shots on goal per game of any player in the world in the last two decades. Messi had 6.1 shots per game, while Ronaldo had 6.8 shots/game (Mazurek, 2018). Note that minutes played, goals scored, assists scored, and titles won have a significant positive influence on a football player market values (Herberger & Wedlich, 2017). According to Transfermarkt database, Ronaldo reached the highest market value when he played for Real Madrid C.F. at €120.00 million in 2014, 2015 and 2018 at ages of 29, 30 and 32 years old (Transfermarkt, 2022b) while Messi achieved the peak of market value at €180.00 million in 2018 at age of 30 years when he played for FC Barcelona (Transfermarkt, 2022a).

In another study, Mazurek (2018) tried to identify a figure that bears most resemblance to Messi. His finding was surprising that Philippe Coutinho, a Brazilian footballer that played with him at Barcelona during 2018-2021 seasons was identified as the most resemblance to Messi. In addition, the study also identified Pierre-Emerick Aubameyang, now playing as captain of the Gabon national team, as the most similar figure with Ronaldo using a dataset from the 2017/2018 season. Unfortunately, both Coutinho and Aubameyang were not included in the top 20 players based on data of the 2000/2001-2022/2023 seasons (Grbec et al., 2024).

More complex analysis for determining the best season and career of a footballer can be done using the ModK formula proposed by Salvador *et al.* (2022). However, a lot of factors need to be considered in the formula. It includes individual indicators in national teams and clubs, indicators of games and titles in national teams and clubs, the impact of the athlete on national teams and clubs, and also the usage of points per match (i.e. the average points earned during the presence of the athlete in the team).

Sarmento and Araújo (2020) stated that an expert performance and high achievement of a football player emerge from circumstances resulting from the self-organization of several performers-environmental factors. Ronaldo and Messi are indisputable examples of ideal football players who have fulfilled all requirements (e.g. speed, strength, agility, endurance, mentality, etc.) to be the best among others. It is not surprising that the current study proved that both players had similar consistency in scoring goals in most of their professional careers as football players.

Finally, Clegg and Robinson (2022) argued in their book that the discussion of "Ronaldo or Messi?" for millions of people around the globe is not simply a barroom argument or an affirmation of fandom, but rather a philosophical statement. Indeed, the philosophical debates may have contributed to the commercialization of football, because the two players have likely driven significant commercial interest and revenue for clubs, leagues, and brands. As a result football strengthen its global reach with these two icon players in turn inspiring new young football players attracted by the value and image of global football today, and perhaps motivated to shape the future of football.

There are several limitations of this study that need to be highlighted including the absence of a statistical method that allowed the researchers to compare the CFM reliability curves of Cristiano Ronaldo and Lionel Messi. To overcome this limitation, we utilized confidence interval theory approach to draw conclusions from the CFM analysis. Furthermore, other factors having the potential to affect the performance of the players were also not included in the analysis due to data limitations, such as physical fitness while playing, post-injury conditions, match importance and home/away status. In addition, Ronaldo and Messi did not always play in the same competition. For example, during the research period they only played in the same league when Ronaldo defended Real Madrid C.F. and Messi played for FC Barcelona in the Spanish La Liga for the 2009-10 to 2017-18 seasons. Nevertheless, examining data from the 19-year career of Cristiano Ronaldo as professional football player and 17-year career of Lionel Messi are believed to be able to provide significant information regarding the comparison of the scoring reliabilities of two extraordinary players in the modern football era.